\documentclass[aps,pre,onecolumn,showpacs]{revtex4}
\usepackage{graphicx}
\usepackage{hyperref}
\usepackage{xcolor}
\usepackage{ccaption}
\usepackage{float}
\begin{document}
\title{Nanocurrent oscillator indefinitely powered by a capacitor battery}
\author{Luigi Ragni}
\email{luigi.ragni@unibo.it}
\affiliation{University of Bologna, Piazza, G. Goidanich, 60, 47521 Cesena (FC), Italy}
\date{\today}

\begin{abstract}

Some electrolytic capacitors show dielectric behaviour that can not be entirely explained by the well known long lasting relaxation. Extra charges able to generate a useful conduction current can be detected for an indefinite time. A squarewave oscillator based on MOSFET CMOS technology and requiring less than 2 nW was powered for 80 days at 25 $\ensuremath{^\circ}$C by a 58.2 mF capacitor battery, without voltage decrease during the last 53 days of observation. The battery consisted of three series of 16 parallel, 15 years aged, capacitors with DC capacitance of 10.9 mF. Capacitors so old, stored without voltage application, were affected by degradation and thinning of the alumina layer that could promote tunnelling of the charge. The main purpose of the present study is to stimulate further investigations aimed at confirming or disputing the observed phenomenon and, if necessary, at shedding light on its physical mechanisms.
\end{abstract}


\maketitle

\section{\label{sec:level1}Introduction}

The charging and discharging of an ideal capacitor are ruled out by simple equations derived from the Ohm law, the Kirchhoff rule and the capacitance definition. The current decrease in the time \textit{t} of that capacitor, with capacitance \textit{C}, after the application of a voltage \textit{$V_0$} through a series resistance \textit{R}, is described by the equation $I(t)=\left(V_0/R\right)\textnormal{e}^{\left(-t/RC\right)}$. During the discharge, if we accurately measure the initial voltage, the capacitance, and the value of the applied resistor, we should be able to predict the voltage drop during the time with the same accuracy, but this is not the case for a real capacitor. The more the electrical resistance increases the more the discrepancy is magnified. It is known that a real capacitor retains more charge at the time \textit{t} than the theorised charge. A spectacularization of this behaviour is the resurfacing of some part of the initial charge at the electrodes if left open ($\textit{R}$$\rightarrow$ $\infty$) after the discharge of the capacitor with a short ($\textit{R}$ $\rightarrow$ 0).
This re-emerging charge varies from about 0.01 to 10 $\%$ of the previous applied one and depends on the kind of dielectric. The phenomenon was first described more than 150 years ago when, in 1854, Kohlrausch observed a long lasting relaxation in a Leyden jar \cite{kol1}. Since then, several different terms have been adopted to describe the same phenomenon: dielectric absorption, dielectric relaxation, soakage, voltage retention and so on.
The nature of dielectric relaxation is not yet fully understood. To date, however, the charging current due to a voltage step application was substantially expressed by power or stretched exponential functions assuming the form of the Curie-von Schweidler law $j(t)=at^{-\alpha}$ or $j(t)=b\exp\left[-(t/\tau)^{\beta}\right]$ where the prefactors \textit {a} and \textit {b} are temperature and applied voltage dependent, 0$<$$\alpha$$<$2, 0$<$$\beta$$<$1 (both $\alpha$ and $\beta$ are again more or less temperature dependent), and $\tau$ is a time constant. Several studies were carried out on this subject, aimed at investigating both the physical mechanisms of the phenomenon and more practical possibilities of solving technical problems that the phenomenon involves, jeopardizing the performances of electronic devices.

Several physical models were suggested to describe the dielectric relaxation. In most of these, tunneling and/or thermally activated processes play the role of the main actors: charge transition between two energetic sites, separated by a barrier \cite{poll1,lew1}; charges (e.g. protons) fluctuating between double well potentials \cite{kli1,and1,phi1,jam1}; dipoles that respond to the applied field and mutually interacting \cite{kuh1,kuh2}. Other modelled mechanisms deal with the space charge polarization \cite{wag1,wag2,mar1,mar2}, where the ions mobility depend on the layer blocking thickness and charges can remain attracted by their image for thin layers (more details on these mechanisms can be found in the paper by Kliem \cite{kli2}).
From a more technical point of view, dielectric absorption (relaxation) can be attributed to a complex capacitance (and impedance). Herein, the main capacitance have to be intended as connected in parallel to a network of further capacitances with series resistances forming RC pairs with different time constants \cite{col1,dow1,hyy1,fatt1,ior1}. An electrical short only discharges the main capacitor and, when it is removed, the network RC pairs recharge the capacitor.

It is our opinion that the dielectric relaxation explains only a part of the complex phenomenon of the charge accumulation after the short in a capacitor, or in other words, that there is a superposition of different phenomena.
In 1994, Neagu and Neagu \cite{nea1} showed dielectric behaviour of PET that diverges from the behaviour based on the Curie-von Schweidler law. The Authors stressed that the discharge was characterized by a very slow decay or a current increase over time.
In 2002 and 2007, Lambert et al. \cite{lam1} and Vrublevsky et al. \cite{vru1} respectively showed that charges are embedded in $Al_20_3$ during anodization and give a true permanent polarization of the layer. In their study, Lambert et al. also reported a comprehensive synthesis of the interpretations regarding the origin of these charges: incorporation of electrolytic ionic species during oxide formation \cite{zud1,hic1}; hydration of the oxide \cite{pop1,kem1}; electrons injected and trapped in the layer \cite{hic2,mik1}.
In 2010, D'Abramo \cite{dab1} modelled a spherical vacuum capacitor with the inner sphere coated by semiconductor material able to generate a difference of potential at the electrodes by thermo-charging action at room temperature.
In 2012 an experimental study \cite{rag1} carried out on industrial electrolytic capacitors with $Al_20_3$ as dielectric showed that some samples loaded with a resistor retained an appreciable charge for more than three months. The study demonstrated the conductive nature of this current, that responds to the Ohm laws, its dependence on temperature and leakage current, a characteristic related to the thickness and integrity of the dielectric layer.

When we consider some of these results we have immediately to deal with the second law of the thermodynamics. Several studies have been carried out on the theoretical possibility of its violation: in a Brownian quantum particle strongly interacting with a quantum thermal bath \cite{all1,all2}, in a superconducting loop with weak links \cite{ber1}, in modelled \cite{zap1} and experimental \cite{wei1} superconducting quantum interference devices or, more generally, regarding the different aspects of thermodynamic limits \cite{she1}. The  above-mentioned references are to few to adequately represent the numerous studies on this matter during the past thirty years, and the reader can, for instance, consult the reference collection in the article by D'Abramo \cite{dab1}.

While the modellization and analysis of interfaces and devices that violate the second law formerlty appeared worthy of consideration by the scientific community, at present there is not widely accepted experimental confirmation of this dramatic possibility. One of the crucial obstacle is the quantum scale of the involved energies, often dominated by the noise, hence the need to transpose processes from a quantum or, in any case, a sub-picometric world to an easily measurable and comfortable classical scale.

Herein we consider the exotic properties of electrolytic capacitors characterized by a certain degree of $Al_20_3$ layer dissolution due to ageing that reduces the insulating barrier and could promote tunneling phenomena of charges \cite{kli2,hic3}. Based on previous observations \cite{rag1} regarding this behaviour, we report on a merely experimental application of what may be an easily repeatable embodiment which shows an apparent violation of the second law of thermodynamics.

\section{\label{sec:level2}Materials and methods}
\subsection{\label{sec:level3}The oscillator}

\begin{figure}
\center
\includegraphics{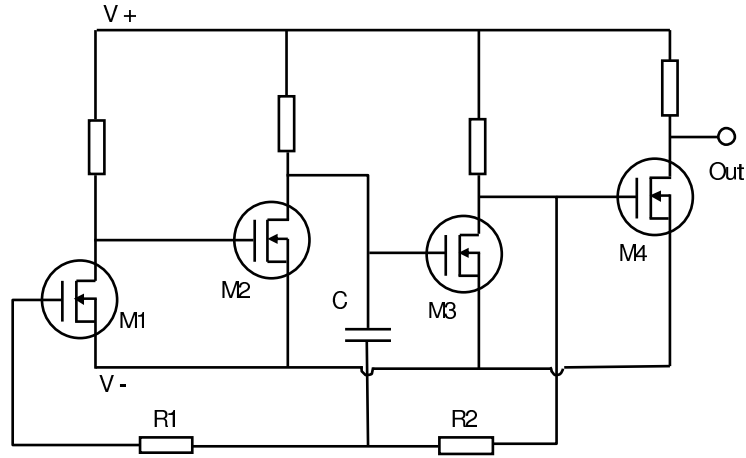}
\caption{\label{fig1:epsart}Schematic of the ultra-low power oscillator \cite{pri1} used in the experiment. (Legend: R1, R2, resistor network and C, capacitor network; M1-4, N-channel Mosfet ALD110804 (M4 has buffer functions).}
\end{figure}

One of the lowest power consuming electronic oscillators was assembled and set up for the present experiment. The device is based on a monolithic quad enhancement mode N-channel Mosfet, matched at the factory by using programmable CMOS technology (ALD110804, Advanced Linear Device, Inc.). The chip is suitable for use at very low operating voltage or very low power, in the order of nanowatt. Although the quad Mosfet used has a precision gate threshold voltage of +0.40V, it can operate in a so called \textit{subthreshold} region where the drain current drops, for example, to 1 nA at Vgate-source = -0.14V. Three Mosfet form a three stage RC oscillator with a feedback resistor, two resistors and a capacitor representing the network. The fourth amplifier acts as a buffer. The circuit, in its original configuration \cite{pri1} (Fig.~\ref{fig1:epsart}), is expected (by simulation) to produce square-wave like oscillations at about 40 Hz, if supplied with 0.14 V. The power required is 1.2 nW. In our application we used a larger capacitance (470 pF) than that the suggested one (100 pF) to obtain a lower oscillation frequency (some resistor values were also slightly changed with respect to the original schematics). After assembling, the circuit was tested to assess its electrical behaviour (voltage, current and oscillation produced).

\subsection{\label{sec:leve4}The capacitor battery}
Forty-eight wet electrolytic capacitors manufactured in June 1997 with rated capacitance of 10 mF, and voltage of 10 V were selected to be a potential current generator able to supply the ultra-low power oscillator at room temperature (25 $\ensuremath{^\circ}$C). It is expected that in so aged capacitors the alumina layer is naturally degraded because of the long storage time without current application that electrochemically regenerates the layer. Scanning electron microscopy was used to compare the anode foil from an aged capacitor with that of an equivalent one, manufactured in December 2011. In a previous study it was found \cite{rag1} that a single capacitor of 3.3 mF can generate a current of about 5 nA with a voltage around 5 mV at the above mentioned temperature.

The capacitors used in the present study are delivered by the vendor with a variable voltage depending on previous charges, storage duration and conditions, accidental short of the terminals, and on their intrinsic characteristics. With the aim of removing the residual charge, the capacitors were conditioned before being connected to the oscillator. Firstly, they were stored at 25 $\ensuremath{^\circ}$C for 72 h and, then, gently discharged by loading the capacitors with a resistor of 991 $\Omega$ for 30 s followed by a resistor of 9.91 $\Omega$ for a further 30 s. During this operation, carried out by an automatic temporized device, the discharge curve was logged and used to measure the voltage before test (BVT) and calculate the DC capacitance (DCC). Other main electrical characteristics, such as the AC capacitance (ACC) at 120 Hz, the leakage current (LC), the dissipation factor ($tg\delta$), and the equivalent series resistance (ESR), were measured to characterize the capacitors. So as not to inject additional charge in the capacitors, these last measurements were conducted on a sample from the same batch of capacitors.
After the discharge by resistors, the terminals were finally shorted (R$\rightarrow$0) and the capacitors stored for 48 h; the short was then removed and the capacitors were stored again for the time necessary to regain a useful charge. All the storage operations were conducted in a PID controlled thermostatic chamber expected to maintain the temperature within 25 $\pm$0.1 $\ensuremath{^\circ}$C, and by placing the capacitors in a grounded metallic container. Although the temperature has been proven to have a high influence on the conduction current \cite{rag1}, no temperature variations were imposed in the present experiment, because the equilibrium of such high capacitance could require very long time periods. Two further capacitors from the same batch, previously submitted to the above mentioned conditioning procedure, were selected to monitor the voltage increase over time. For this purpose an automatic measurement apparatus was assembled and subsequently used for the supply oscillator main test. Based on the data from these two capacitors, the storage was prolonged until the voltage at the terminals left open of the 48 capacitors was expected to be at least 80 mV. Three branches of 16 capacitors in parallel were subsequently formed by choosing the single capacitors in a way that each branch has similar resulting total voltage and standard deviation. The three branches were finally connected in series to obtain a capacitor battery with a voltage of at least 240 mV and a rated capacitance of 53.3 mF.

\subsection{\label{sec:level5}The automatic measurement apparatus}

\begin{figure}[H]
\center
\includegraphics{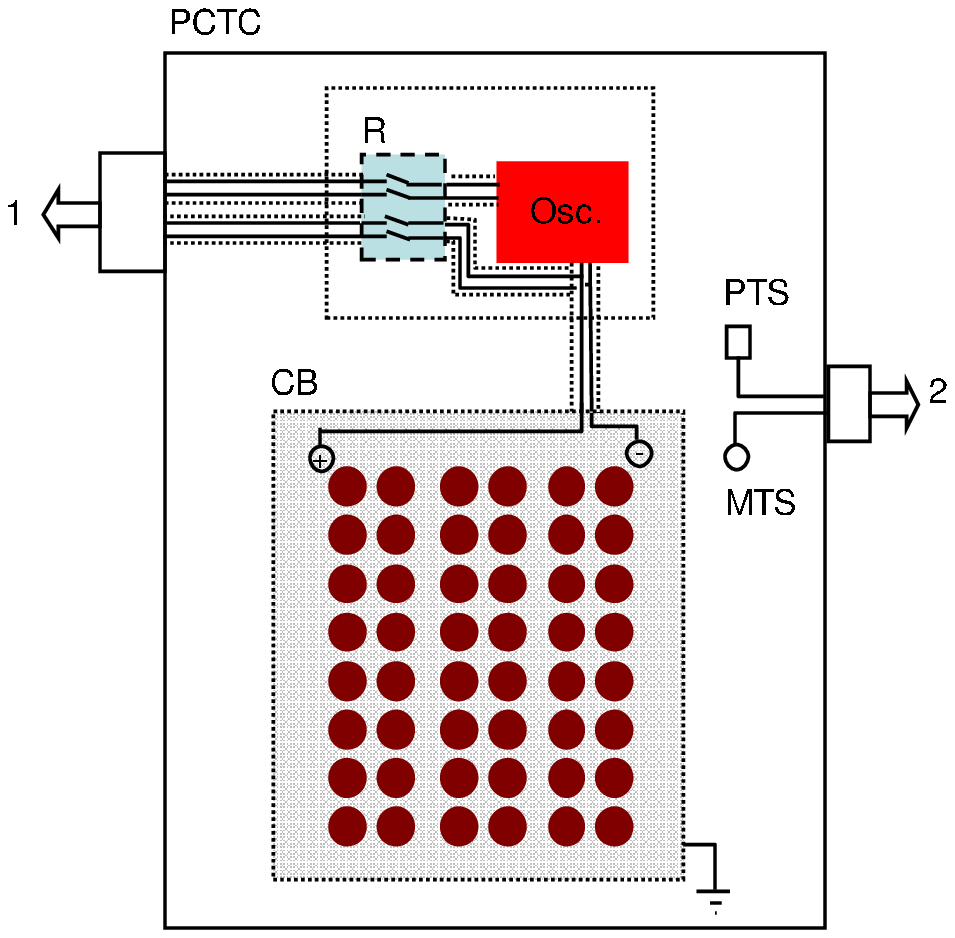}
\caption{\label{fig2:epsart} Layout of the experimental apparatus.
\newline
\footnotesize (Legend: PCTC, PID controlled thermostatic chamber; R, reed relays; Osc, oscillator; CB, capacitor battery; PTS, temperature sensor to control the PID device; MTS, sensor to measure the temperature inside the chamber; 1, outputs to the acquisition board; 2; outputs to the acquisition board and PID controller.)}
\end{figure}

The systems consisted of two metallic shielding and grounded boxes. One of these housed the oscillator and the reed relays (further shielded) to temporize the connections from the supply capacitor battery and the oscillator outputs to the digitizing board. The capacitor battery permanently supplied the oscillator through a shielded bipolar grounded cable. The electrical lines from the capacitors and from the oscillator to the digitizer were maintained as short as possible and interrupted by relays so that they were isolated from the measurement system for most of the time. In the main test (capacitor battery that supplies the oscillator), the relays connected the lines for 3 s each day and the voltage of the oscillator output and capacitor battery were logged for 2 s. In the preliminary test, aimed at assessing the regain of capacitor charge, the measurement were logged for 1 s each hour. Acquisition and relay activation were performed by Labview 8.2 software (National Instrument) and by using an acquisition board (6024E, 12 bit, NI) with high input impedance. Sampling frequency was set to 500 Hz. A further line was provided to monitor the temperature inside the thermostatic chamber. A layout of the described system is sketched in Fig.~\ref{fig2:epsart}.
The measurements of the voltage battery were carried for long enough to check a possible suspension of its decrease. A test to exclude the injection of charges through the lines (coaxial cables) connecting the oscillator and the battery to the acquisition board was subsequently carried out. The lines were disconnected and the relays switched off for 10 days, the voltage of the battery was then measured.

\section{\label{sec:level6}Results}
\subsection{\label{sec:level7}Oscillator characterization}

\begin{figure}
\center
\includegraphics{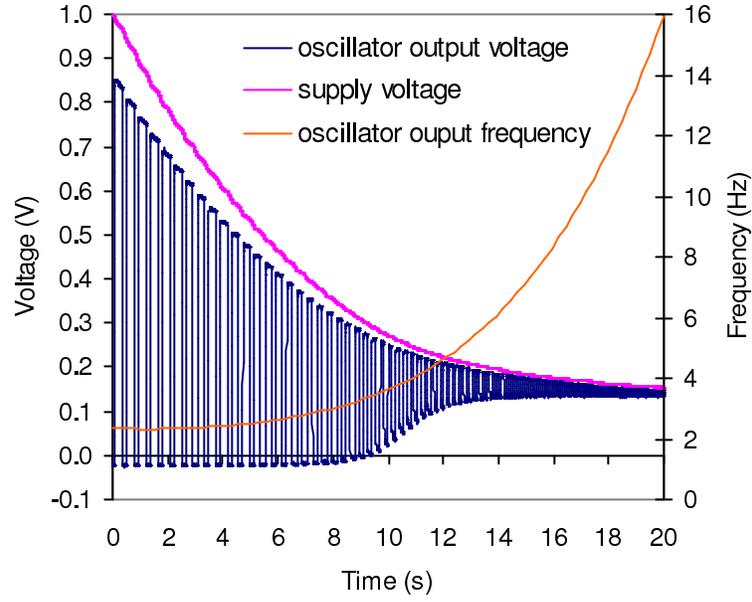}
\caption{\label{fig3:epsart} Dependence of the supply voltage, given by a 0.78 $\mu$F capacitor discharge, on the assembled oscillator output.}
\end{figure}

\begin{figure}[H]
\center
\includegraphics{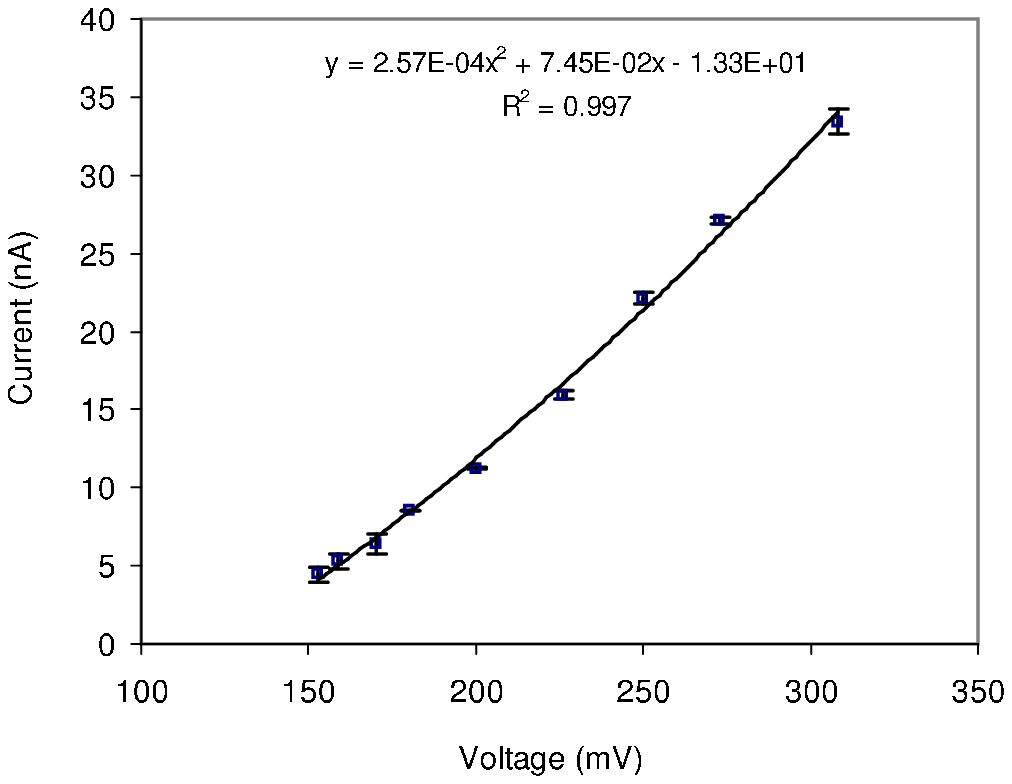}
\caption{\label{fig4:epsart} Correlation between supply voltage and current of the assembled oscillator.}
{Note: Error bars show the standard deviation.}
\end{figure}

The electrical behaviour of the assembled oscillator circuit is summarized in Fig.~\ref{fig3:epsart} and Fig.~\ref{fig4:epsart}. Fig.~\ref{fig3:epsart} shows the dependence of the oscillation amplitude and frequency on the supply voltage, from 1 V to 0.15V, furnished by the discharge of a capacitor with a capacitance of 0.78 $\mu$F. Fig.~\ref{fig4:epsart} shows the correlation between supply voltage and current.

\subsection{\label{sec:level8}Capacitor characterization}

Table \ref{table:nonlin} shows the main electrical and geometrical characteristics of the kind of capacitors used in the present experiment.

\begin{table}[hc]
\caption{Measured main electrical and geometrical characteristics of the capacitor.
\newline
\footnotesize {Legend: DCC, direct current capacitance; ACC, alternating current capacitance; VBT, voltage before test; LC, leakage current; tang $\delta$ , dissipation factor; ESR, equivalent series resistance. EFA, capacitor element foil area (anode plus cathode foils); WED, wound element diameter; WN, wounds number; AT, anode, CT cathode, and PT, paper thickness.}}
\centering
\begin{tabular}{c c c c c c}
\hline\hline
DCC ($\mu$F) & ACC ($\mu$F) & VBT (V) & LC ($\mu$A) & tang$\delta$ & ESR (m$\Omega$) \\ [0.5ex]
10916$\pm$34 & 9975$\pm$59 & 0.497$\pm$0.037 & 36$\pm$3.7 & 0.142$\pm$0.004 & 18.9$\pm$0.6 \\
\hline
EFA ($cm^2$) & WED (mm) & VN & AT ($\mu$m) & CT ($\mu$m) & PT ($\mu$m) \\ [0.5ex]
487.8$\pm$0.2 & 15.3$\pm$0.2 & 26 & 67.49$\pm$0.9 & 47.3$\pm$0.9 & 26.9$\pm$2.1 \\ [1ex]
\hline\hline
\end{tabular}
\label{table:nonlin}
\end{table}

\begin{figure}[H]
\center
\includegraphics{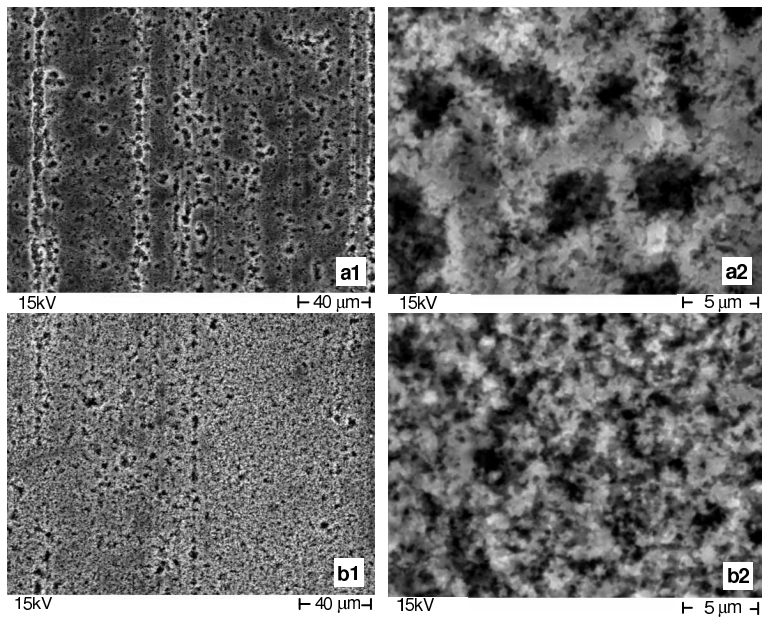}
\caption{\label{fig5:epsart} SEM images of anode foil from (a) the batch of capacitors used in the present study and (b) from a recently manufactured capacitor (thanks to Roberta Randi).}
\end{figure}

Fig.~\ref{fig5:epsart} shows SEM images of the anode foil from the capacitors used in the present experiment (a) and in a newly manufactured one (b), for comparison.

As one can see, in the aged capacitor the anode foil is characterized by spread pores due to the dissolution processes that occur in electrolytic capacitors unconnected to a source voltage for a long time.

\begin{figure}[H]
\center
\includegraphics{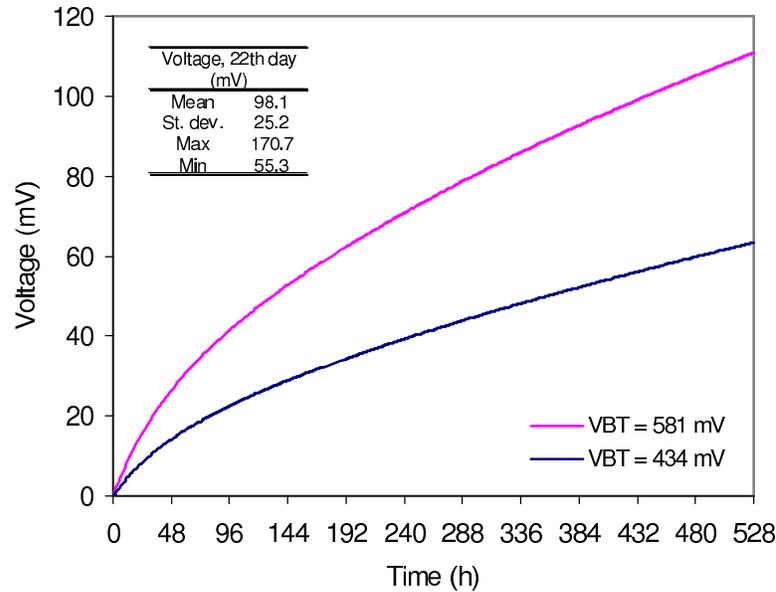}
\caption{\label{fig6:epsart} Voltage re-emerging of two indicator capacitors with different voltage before test (VBT) and with terminals left open during 528 h (22 days) of storage at 25 $\ensuremath{^\circ}$C.
\newline
\footnotesize {In the box: mean, standard deviation, maximum and minimum values of all the 48 capacitors subsequently used to form the capacitor battery.}}
\end{figure}

Fig.~\ref{fig6:epsart} shows the re-emerging voltage of two selected capacitors with terminals left open, after discharge with resistors and 48 h of short, during 528 h (22 days) of storage at 25 $\ensuremath{^\circ}$C. The data from this test was used to estimate the time necessary to form a battery of three branches in series of 16 capacitors in parallel to obtain a 240 mV battery.

\begin{figure}[H]
\center\includegraphics{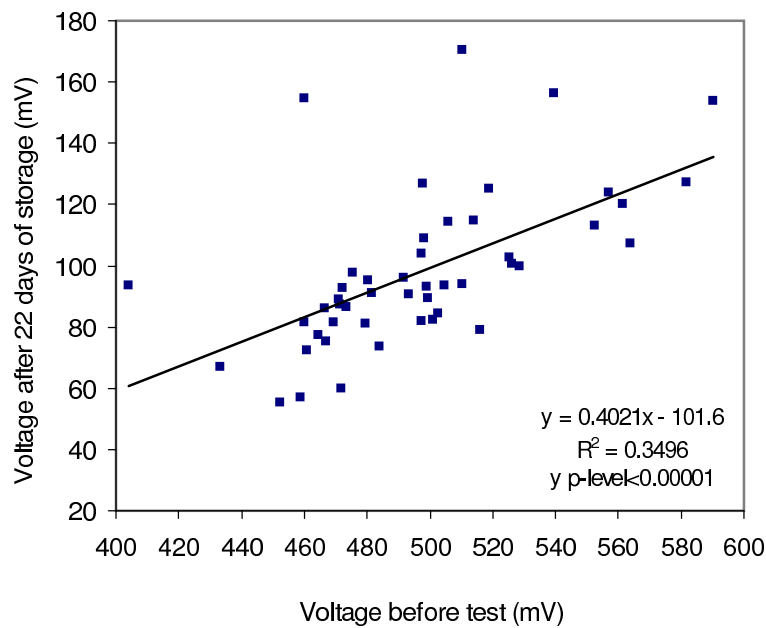}
\caption{\label{fig7:epsart} Correlations between the voltage of the 48 capacitors before test and 528 h (22 days) after the short was removed, at 25 $\ensuremath{^\circ}$C.}
\end{figure}

Fig.~\ref{fig7:epsart} shows the correlation of the re-emerged voltage after 528 h (22 days) from the 48 h short removal and the voltage before test.

The voltage before test, which is the voltage of the capacitors on delivery from the manufacturer, is influenced by previous storage conditions, polarization and accidental shorts. Despite this, the re-emerged voltage due to charge accumulation was strongly dependent on it (p-level $\textless$ 0.0001).
After 528 h of storage the capacitor mean voltage was 98.1  $\pm$25.2 mV. From this sample, three sub-samples of capacitors were selected to form three branches with voltage of 98.5  $\pm$25.8 mV, 97.7  $\pm$24.1 mV, and 98.2  $\pm$25.7 mV, connected subsequently in series to give a voltage of 294.3 mV at the terminals of the resulting capacitor (capacitance = 58.2 mF), more than that necessary. 

\subsection{\label{sec:level9}Oscillator power supply and output}

\begin{figure}[H]
\center
\includegraphics{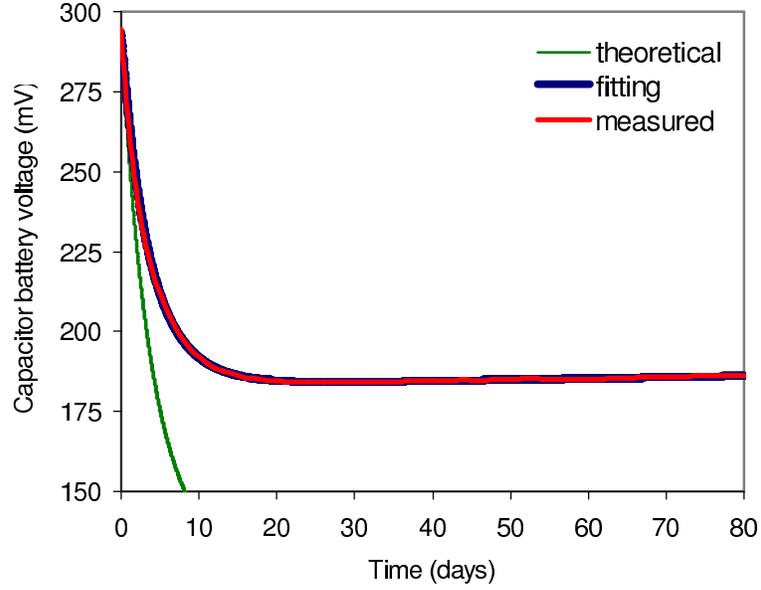}
\caption{\label{fig8:epsart} Voltage of the capacitor battery during 80 days.}
\end{figure}

The measured voltage at the capacitor battery over time after the oscillator switch on is shown in Fig.~\ref{fig8:epsart}.
The voltage continuously dropped from about 294 mV to 184 mV in 27 days following the trend of a function like $y=A\exp\left\{B\left[\exp\left(Cx\right)\right]\right\}$ ($R^2$ = 0.9997) with \textit{A} = 0.1840; \textit{B} = 0.4674; \textit{C} = -0.2429 (y in V, and x in days). During this period the oscillator rms voltage output decreased from 182 mV (pk-pk = 276 mV) to 147 mV (pk-pk = 46 mV) and the frequency increased from 3.27 Hz to 7.25 Hz. In the subsequent 53 days, the capacitor battery voltage remained roughly constant, or better, weakly increased to 186 mV. The function describing this last period can be represented by: y = 3.89 $\times$ $10^{-5}$ x + 0.183. The calculated (fitting) trend from the above mentioned functions and the completely theoretical one (without any considered extra-charge) are shown in the same Fig. 8. The ideal voltage of the capacitor battery should drops to 186 mV in 4.2 days.
To estimate the amount of extra-charge in the capacitors a simple mathematical approximation was applied. Firstly the theoretical, voltage drop of an ideal capacitor was calculated every 0.01 days (864 s), step by step, through the experimental initial voltage (from the above mentioned functions) according to the following equation:

\begin{equation}
	V_t=V_e\left\{\textnormal{e}^{-\Delta t/\left[\left(V_e/I_e\right)C\right]}\right\}
\end{equation}

where: \textit{$V_t$} is the theoretical, ideal final voltage in V; \textit{$V_e$} is the experimental, initial voltage in V, \textit{$\Delta t$} is the time between steps (864 s), in s, \textit{$I_e$} is the experimental, initial current, in A; \textit{C} is the capacitance, in F.

The step by step extra voltage, \textit {$\Delta V$}, was then calculated as:
\begin{equation}
	\Delta V = V_{e+1}- V_t
\end{equation}

where \textit{$V_{e+1}$} is the experimental initial voltage of the step subsequent to \textit{$V_e$}.

The \textit {$\Delta V$} variations were very small during the 80 days of storage: mean = 0.137 mV; standard deviation = $\pm0.002$ mV; min = 0.123 mV; max = 0.141 mV, so that the {$\Delta V$} can be considered roughly constant and corresponding to 159 nV/s or 9.23 nC/s. 

A simplified model can now be constructed according to:

\begin{equation}\label{Eq3}
	V_{m+1}=V_m\left\{\textnormal{e}^{-\Delta t/\left[\left(V_m/I_m\right)C\right]}\right\}+\Delta V_k
\end{equation}

where \textit{$V_{m+1}$} is the final modelled voltage; Vm is the initial modelled voltage of each step (the first \textit{$V_m$} value is coincident with the voltage of the battery at the beginning of the test); \textit{$\Delta$$V_k$} is a constant value of \textit{$\Delta V$}.

\begin{figure}[H]
\center
\includegraphics{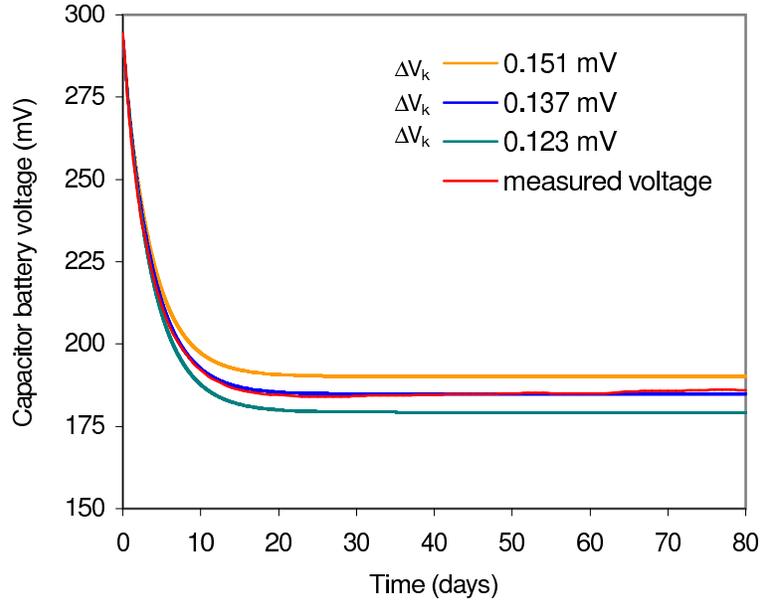}
\caption{\label{fig9:epsart} Behaviour of the model in eq. \ref{Eq3} with different $\Delta$$V_k$ values, and measured voltage.}
\end{figure}

The behaviour of the model with respect to the experimental data is shown in Fig.~\ref{fig9:epsart} for three different \textit{$\Delta$$V_k$} values ($\pm10$$\%$ with respect to the mean value).

In the explored time period, apart from the lack in the light voltage increase description, the model fits the observations quite well. It is quite sensitive to the constant extra voltage value: a variation of 10 $\%$ in $\Delta$$V_k$ involves a diference between the modelled and the measured values up to 3 $\%$.

\begin{figure}[H]
\center
\includegraphics{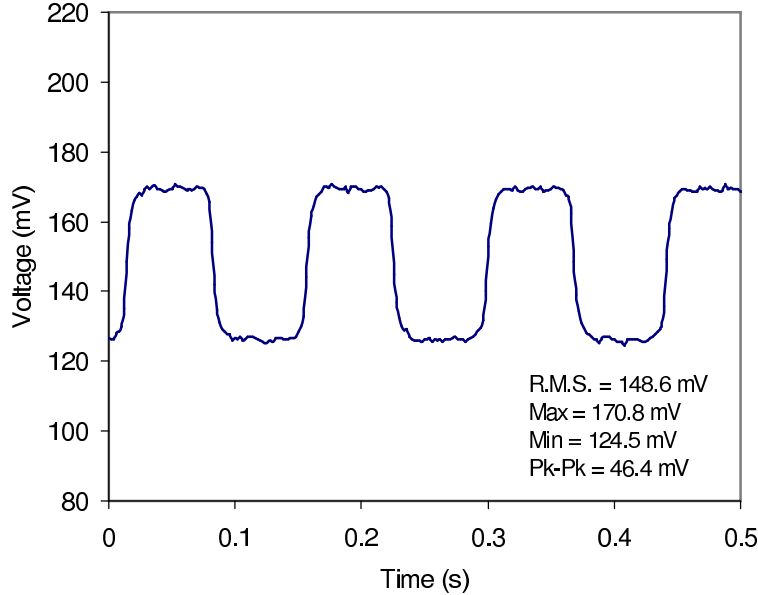}
\caption{\label{fig10:epsart} Oscillator output ($80^{th}$ day) with 186 mV at the capacitor battery.}
\end{figure}

The waveform of the oscillator after 80 days is depicted in Fig.~\ref{fig10:epsart}.
The squarewave was well shaped, with a dominant oscillation frequency of about 7.2 Hz.
The signal appeared weakly affected by noise: an appreciable result if we consider the low voltage level and, above all, the very high impedance of the output.
During the 80 days the mean temperature was 24.99 $\ensuremath{^\circ}$C with a standard deviation of 0.08 $\ensuremath{^\circ}$C. The maximum and minimum logged temperatures were 25.38 $\ensuremath{^\circ}$C and 24.59 $\ensuremath{^\circ}$C, respectively.
Finally, after 10 days of testing without connections from capacitor battery and oscillator output to the acquisition board, no appreciable variation of the voltage was detected, with respect to the logged value at $80^{th}$ day.

\section{\label{sec:level10}Conclusions}

A 58 mF electrolytic capacitor battery appeared able to supply a 1.8 nW squarewave oscillator for 80 days at 25$\ensuremath{^\circ}$C, after being fully discharged, and a usable amount of charge re-emerged without any applied external source. From the $53^{rd}$ day the voltage remained roughly constant if not weakly increasing. The assembled battery consisted of a series of three branches, each of which formed by 16 parallel capacitors. The capacitors used were aged about 15 years without current application: a condition sufficient to degrade and make the alumina dielectric layer thin. The present experiment seems to confirm the results of a previous study carried out with 3.3 mF electrolytic capacitors loaded with a resistor and stored at the same temperature.
On the basis of this evidence it seems possible to indefinitely power a device in the framework of an apparent violation of the second law of thermodynamics. Before jumping to erroneous, rejectable conclusions some considerations have to be made. We do not know what is the prevalent physical mechanism involved in this behaviour; we can only speculate that it could be due to, or at least influenced by the following phenomena: charge tunnelling due to the reduced thickness in the alumina layer and associated with permanent embedding of charges in the layer; thermoelectric or thermionic effects within temperature oscillations of a few tenths of Celsius degree; chemical reactions; background radiation. Again, we can not say anything about the evolution of the present experiment, other than that it can be observed (for some time) at the webpage \url{http://oscillator.polocesena.unibo.it/oscillator.html}. On the other hand we think that this phenomenon deserves attention. \textit{In primis}, we hope that the experiment will be replicated by independent laboratories, better if in more severe test conditions, such as with temperature variations in the order of 1/100 of Celsius degree and with background radiation shielding.

\end{document}